\documentclass[journal]{IEEEtran}
\usepackage{ifpdf}
\usepackage{url}
\usepackage{amsmath,amssymb,amsfonts}
\usepackage{graphicx,cite,epsfig}
\usepackage{algorithm}
\usepackage{algorithmic}
\usepackage{textcomp}
\usepackage{xcolor}
\usepackage{graphicx}
\usepackage{subfigure}
\usepackage{siunitx}
\usepackage{booktabs}
\usepackage{multirow}
\usepackage{float}
\usepackage{caption}
\usepackage{threeparttable}
\usepackage{diagbox}
\usepackage{bm}
\usepackage{color}
\usepackage[colorlinks,linkcolor=blue]{hyperref}

\newcommand{\ieno}{\textit{i}.\textit{e}.}

\ifCLASSINFOpdf

\else

\fi
\begin{document}
%
\title{GraphIQA: Learning Distortion Graph\\
Representations for Blind Image Quality Assessment}
%
%
%

\newcommand{\ssm}{\textcolor{red}}
\newcommand{\yt}{\textcolor{blue}}

\author{Simeng Sun$^*$, Tao Yu$^*$, Jiahua Xu, Wei Zhou and
	Zhibo Chen,~\IEEEmembership{Senior Member,~IEEE,}
\thanks{* Equal contribution.}
\thanks{
Simeng Sun, Tao Yu, Jiahua Xu, Wei Zhou and Zhibo Chen are with the Department of Electronic Engineer and Information Science, University of Science and Technology of China, Hefei, Anhui, 230026, China (e-mail: smsun20@mail.ustc.edu.cn; yutao666@mail.ustc.edu.cn; skylarxu@tencent.com; weichou@mail.ustc.edu.cn; chenzhibo@ustc.edu.cn). Corresponding Author: Zhibo Chen.
}
}

\markboth{Journal of \LaTeX\ Class Files,~Vol.~14, No.~8, March~2021}%
{Shell \MakeLowercase{\textit{et al.}}: Bare Demo of IEEEtran.cls for IEEE Journals}
%



\maketitle

\begin{abstract}


A good distortion representation is crucial for the success of deep blind image quality assessment (BIQA). However, most previous methods do not effectively model the relationship between distortions or the distribution of samples with same distortion type but different distortion levels. In this work, we start from the analysis of the relationship between perceptual image quality and distortion-related factors, such as distortion types and levels. Then, we propose a Distortion Graph Representation (DGR) learning framework for IQA, named GraphIQA, in which each distortion is represented as a graph, \ieno, DGR. One can distinguish distortion types by learning the contrast relationship between these different DGRs, and infer the ranking distribution of samples from different levels in a DGR.
Specifically, we develop two sub-networks to learn the DGRs: a) Type Discrimination Network (TDN) that aims to embed DGR into a compact code for better discriminating distortion types and learning the relationship between types; b) Fuzzy Prediction Network (FPN) that aims to extract the distributional characteristics of the samples in a DGR and predicts fuzzy degrees based on a Gaussian prior. Experiments show that our GraphIQA achieves the state-of-the-art performance on many benchmark datasets of both synthetic and authentic distortions. The code is available at \href{http://staff.ustc.edu.cn/~chenzhibo/resources/2021/GraphIQA.html}{http://staff.ustc.edu.cn/~chenzhibo/resources/2021/GraphIQA.html}.

\end{abstract}

\begin{IEEEkeywords}
blind image quality assessment, graph representation learning, and pre-training.
\end{IEEEkeywords}

%
\IEEEpeerreviewmaketitle

\section{Introduction}
\IEEEPARstart{W}{ith} the rapid development of social networks, a massive amount of digital images have been produced. They could be distorted in any stage of the whole media technical chain, from acquisition, processing, compression to transmission and consumption. Therefore, a reliable image quality assessment (IQA) metric is critical for measuring multimedia model results and guiding its optimization.  





Within the scope of IQA, no-reference or blind image quality assessment (NRIQA or BIQA) has drawn much attention since the references are often not available in many real-world applications. Meanwhile, learning-based BIQA methods perform well thanks to the powerful fitting capacity of deep neural networks~\cite{ZBChen20@stereoscopic,su2020blindly,zhu2020metaiqa,zhang2018blind,zhou2019dual,kang2014convolutional,ma2018end, golestaneh2020no,chen2017blind}. 

A good representation could help the training of the target task~\cite{RepresentationLearning}. Particularly, in situations where labeled data are hard to reach, the representations obtained by unsupervised~\cite{UnsupervisedRepresentation, he2020momentum, DINO} or semi-supervised~\cite{BERT,liu2017rankiqa,zhang2018blind} learning can serve as auxiliary information to solve the supervised learning tasks.
Recently, the research on representation learning has helped to make breakthroughs in various fields~\cite{RepresentationLearning,GraphRepresentation,BERT,UnsupervisedRepresentation}.
In IQA, the improvement of performance and model generalization ability is also inseparable from the efficient representations of distorted data~\cite{kang2014convolutional,ma2018end, golestaneh2020no, zhang2018blind, xu2016multi,liu2017rankiqa}.

Many methods improve IQA model performance by learning a good representation of distortion, so as to better serve the quality score regression.
One common approach is introducing an auxiliary distortion classification task in latent space to enforce the feature representations to be discriminative to distortion types~\cite{kang2015simultaneous,ma2018end,golestaneh2020no}, which is one of the important factors affecting image perceptual quality.
Although such type classification task can assist in IQA tasks, the representations obtained by these methods may suffer from at least two issues: 
1) they cannot distinguish the level of distortion, which is also an important factor in image perceptual quality;
2) they are not robust when being adapted to the IQA task for authentic distortion due to the uncertainty of distortion type and non-homogeneity of the authentic distortions.
To address the first issue, Zhang \textit{et al.}~\cite{zhang2018blind} propose to employ an extra distortion-level classification task.
However, they ignore intrinsic distribution properties among distortion levels.
For example, assuming a scene where there are three images with distortion level-$1$, $2$ and $5$ respectively, the methods based on classification task fail to model their ranking relationship. As the levels are treated as independent categories, the ranking relationship, where level-$1$ samples are more similar to level-$2$ samples than level-$5$ samples, can't be discriminated.
Xu \textit{et al.}~\cite{xu2016multi} address the issue by designing a rank model for each distortion to learn the ranking relationship among levels. This method cannot efficiently handle unseen distortion types as there is no corresponding rank model for this type. Then Liu \textit{et al.}~\cite{liu2017rankiqa} propose a siamese network to learn to rank two images sampled from the same distortion. However, it ignores modeling the distortion type.
For the second issue, Zhang \textit{et al.}~\cite{zhang2018blind} attempt to directly perform bilinear pooling of the synthetic and authentic feature sets to achieve better performance on the two kinds of distorted data simultaneously. However, in this method, two pre-trained networks are required to handle synthetic and authentic distortions at the same time. Besides, the computational complexity of bilinear pooling for fusing the two features is also high.


In this work, we model the relationship between distortion type and distortion level as a hierarchical model based on our observations (details are described in Section~\ref{subsec:motivation}) and the conclusions in the mentioned work~\cite{xu2016multi,liu2017rankiqa, zhang2018blind}. That is, learning to rank the samples from specific distortion types and discriminate their levels are beneficial for obtaining better representations for IQA task. Therefore we introduce graph representation, which is suitable for modeling the hierarchical structure when giving proper definitions of node and edge. In detail, each graph itself is used to represent a particular distortion type, while its node distribution in a specific graph is used to represent different distortion levels. In addition, in order to make the learned representations robust across distortion types (including synthetic distortion, authentic distortion, and multiple distortion), we propose to learn the relationships between distortion types by drawing on metric learning methods. 
Overall, the proposed method is a two-stage method. In the pre-training stage, we explore modeling distortion types and levels with a single high-efficient model and learn better distortion representations from distortion contrast relationships and their internal distributions. In the finetuning stage, the learned representations of distortion are used to assist IQA task on target dataset.
To this end, we propose a novel BIQA framework that integrates graph representation learning, dubbed GraphIQA.
\begin{figure}
    \centering
    
    \includegraphics[width=0.8\linewidth]{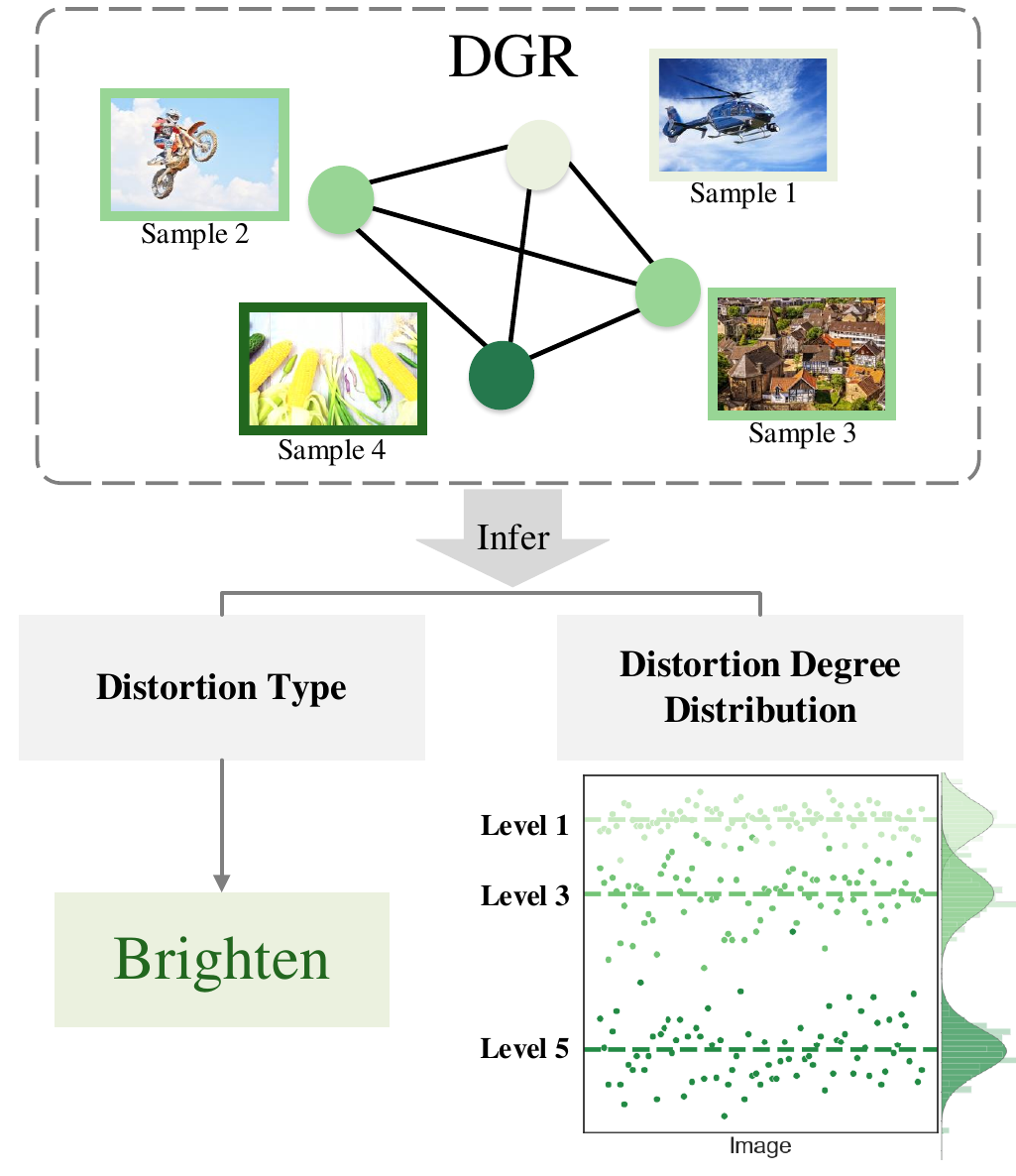}
    \caption{The core idea of GraphIQA. We develop DGR to represent each distortion.
    The DGR can be utilized to infer distortion type and level based on its internal structure. 
    The DGR learning also considers the rating deviation of image content for better prediction.
    The learned DGRs with plentiful distortion prior can help improve IQA accuracy.\label{fig:firstpic}}

\end{figure}

The proposed GraphIQA model is trained to build the distortion graph representation (DGR) for each specific distortion. In each DGR, the nodes represent the feature of samples and the edges illustrate their correlation. The core idea of GraphIQA is shown in Fig. \ref{fig:firstpic}, where DGR is constructed from two aspects: (a) distinguishing the distortion type by contrasting the DGRs of different distortions; (b) predicting the most likely distortion level of a distorted image according to the internal topological relationship in each DGR. 
To achieve these two goals, we correspondingly design Type Discrimination Network (TDN) and Fuzzy Prediction Network (FPN) to learn the DGRs respectively. In detail, the TDN encodes DGR to a low-dimensional code to distinguish distortion types by aggregating the global information of nodes and the relationship between them. 
Specifically, it discriminates each distortion type and learns a robust representation of relationship between types by enforcing a triplet loss~\cite{schroff2015facenet} on the top of the extracted code. 
Then the FPN extracts the distributional characteristics of the samples in DGR and predicts fuzzy levels based on a Gaussian prior considering the subjective quality ratings are often biased by image content. 
The visualization experiments show that the learned DGRs can model the relationship between perceptual image quality and distortion-related factors. Benefit from the DGR GraphIQA achieves the state-of-the-art performance on the most of the typical synthetic distorted IQA datasets (\textit{e.g.} after finetuning, LIVE~\cite{sheikh2006statistical} and CSIQ~\cite{larson2010most}). The experiments also demonstrate that GraphIQA can be migrated to multiply distorted data (\textit{e.g.}, LIVEMD~\cite{jayaraman2012objective}) and authentic distorted data (\textit{e.g.}, KonIQ-10k~\cite{lin2018koniq} or (LIVEC)~\cite{ghadiyaram2015massive})), and obtain better performance.
Note that the data used for pre-training DGRs is within easy reach as it only requires synthetic distorted data and their labels of distortion type and level for weakly supervised training. Our contributions can be summarized as follows:

\begin{itemize}
    \item We investigate the inherent relationship between distortion-related factors and their effects on perceptual quality and propose an effective Distortion Graph Representation (DGR) learning framework dubbed GraphIQA for general-purpose BIQA task.

    \item To encourage better graph representation learning for the relationship modeling of distortion-related factors in GraphIQA, we well design a Type Discrimination Network (TDN) and a Fuzzy Prediction Network (FPN) to learn the proposed DGR.
    \item 
    The proposed DGR can be conveniently applied in most downstream IQA tasks including IQA of synthetic distortion, authentic distortion and multiply distortion, and help GraphIQA achieve the state-of-the-art performance.
\end{itemize}

The rest of the manuscript is arranged as follows. Recent progress on blind image quality assessment and graph representation learning are introduced in Section~\ref{relate works}. The motivation and details of the proposed GraphIQA framework are introduced in Section~\ref{method}, and corresponding experiments are illustrated in Section~\ref{exp}. We conclude this paper in the last section.

\section{Related Work}
\label{relate works}

\subsection{Blind Image Quality Assessment}
\label{sub:nriqa}

Blind Image Quality Assessment(BIQA) can be categorized into distortion-specific methods~\cite{li2013referenceless,li2015no,liu2009no,lin2019kadid} and general-purpose algorithms~\cite{WZhou20@tensor,saad2012blind,mittal2012no,zhang2015feature,xue2013learning,xue2014blind,ke@Using,ye2012unsupervised,xu2016blind,ghadiyaram2017perceptual,fang2014no,LKShi20@no,chen2017full,jiang2017optimizing, guan2017visual}. The distortion-specific BIQA methods are favored for their higher accuracy and robustness, when distortion types or distortion process is already known. However, their application scope is limited, as the authentic distortion dataset is mixed with complex distortions and the type of distortion is not clearly specified~\cite{ghadiyaram2015massive,jayaraman2012objective}. Therefore, the research on general-purpose methods has become particularly important and received extensive attention recently. Natural scene statistics (NSS) is one of the powerful tools for general-purpose BIQA, as quality degradations can cause deviation from the originally statistical properties of natural scene images~\cite{mittal2012no,saad2012blind,mittal2012making,zhang2015feature,li2016blind,yan2019naturalness, ke@Using}. For example, Saad \textit{et al.}~\cite{saad2012blind} leverage the statistics of local DCT coefficients as the feature for image quality assessment, while Moorthy \textit{et al.}~\cite{moorthy2011blind} leverage the feature obtained from the wavelet transform. To simplify the process of feature extraction, Mittal \textit{et al.}~\cite{mittal2012no} propose the method using the NSS in the spatial domain directly. And Zhang \textit{et al.}~\cite{zhang2015feature} leverage not only the statistics of the mean subtracted contrast normalized coefficients, but also the statistics of gradients.
Recently, benefit from its ability to efficiently and adaptively extract distortion-aware features, the deep learning-based general-purpose BIQA methods have drawn considerable attention\cite{kim2017deep,deng2009imagenet,talebi2018nima,zeng2017probabilistic,kang2015simultaneous,ma2018end,zhang2018blind,liu2017rankiqa,xu2016multi,yang2020ttl}. Kim \textit{et al.}~\cite{kim2017deep} propose an efficient approach and prove that using backbone pre-trained on large classification dataset ImageNet~\cite{deng2009imagenet} can improve the performance of IQA. Based on this, Talebi \textit{et al.}~\cite{talebi2018nima} propose a DCNNs-based model to predict the perceptual distribution of IQA scores instead of the mean value. Similarly, Zeng \textit{et al.}~\cite{zeng2017probabilistic} propose the probabilistic quality representation to describe the image subjective score distribution. Noticing that, in synthetic distortion data, effectively utilizing distortion-related information is a common approach to help the learning of representation for IQA task, Kang \textit{et al.}~\cite{kang2015simultaneous} introduce a compact multi-task network into IQA in which type identification task and IQA share all the internal structure.
Ma \textit{et al.}~\cite{ma2018end} introduce a two-stage training strategy, where a distortion type identification sub-network is first trained, and then a sub-network for IQA task is added. Though multi-task related methods have brought progress in IQA, this is difficult to be utilized on authentically distorted datasets, as the representation learned by type classification task can't handle totally unseen distortion types.
For better performance on both synthetic and authentic distortion, Zhang \textit{et al.}~\cite{zhang2018blind} combine two sub-networks, one of which is trained on type classification task aimed to extract features to represent synthetic distortion, the other of which is ImageNet pre-trained model aimed to extract semantic features. Then two kinds of features are fused by bilinear pooling to predict the subjective quality score. Another way of pre-training strategy is learning from rankings. Xu \textit{et al.}~\cite{liu2017rankiqa} train a Siamese Network to rank images in terms of image quality by using synthetically generated distortions for which relative image quality is known. And to learn type-specific ranking rules, Xu \textit{et al.}~\cite{xu2016multi} train a ranking model for each clustering of distortion type. The former only considers the rankings between samples while ignoring another important distortion-related factor, \ieno, distortion type. The latter performs rank learning for each type, requiring training the same number of branching networks as the distortion type, which leads to a great increase in network complexity when types get more.

Here, we design a novel framework to learn better representation for both relationship between distortion types and distortion levels (\ieno, learning to discriminate then rank). The main idea can be concluded as two aspects: one is modeling the distortion-related factors as graph model instead of plane model realized by classification task, the other is learning the relationship between distortion types for better generalization to unseen distortion types. 

\subsection{Graph Representation Learning}
\label{sub:representation learning}

A graph can represent data that are generated from non-Euclidean domains with relationships and inter-dependency between data~\cite{GraphRepresentation}. The challenge in graph representation learning is finding a way to properly represent/encode the graph structure so that it can be easily integrated into the machine learning model. Most of the traditional methods are based on hand-crafted features, such as statistics or kernel functions. Recently, encouraged by the success of CNNs in the computer vision field, a large number of methods that are based on automatically learned low-dimensional embeddings to encode the structure of graphs have been developed. Having the ability of neighborhood aggregation, Graph convolutional networks (GCNs) have been successfully applied to 
many tasks~\cite{xu2020blind,berg2017graph,yan2019convolutional,fu2020bayesian,xu2020blind}. Graph attention network (GAT)~\cite{velivckovic2017graph} further integrates masked self-attention mechanism in GCN. Different from the aggregation method of weighted sum in GCN, Hamiton \textit{et al.}~\cite{hamilton2017inductive} propose GraphSAGE, which introduced an inductive learning mode. By training the model to aggregate neighbor nodes using max-pooling and LSTMs~\cite{hochreiter1997long}, GraphSAGE is extended to inductive learning task, so that it can achieve the generalization for unknown nodes. However, the mentioned methods are based on neighborhood aggregation resulting in the shallow representation of graph, which prevents the model from obtaining adequate global information. Therefore, Hu \textit{et al.}~\cite{hu2019hierarchical} propose hierarchical graph convolutional network (H-GCN) with a graph pooling mechanism to solve the above problem, showing great improvement. In this paper, we are inspired on that the effect of distortion on the perceptual image quality is not only manifested in the characteristics of distortion itself, but also the distribution of samples at different levels under that distortion. There is a hierarchical relationship between distortion type and level, which is appropriate to be modeled as graph. Therefore, we introduce graphs to efficiently represent various distortions, which will be used to help the representation learning of distortion than improve the performance on IQA task.

\begin{figure}[t]
\centering
    \centering
    
    \subfigure[]{
    \includegraphics[width=1\linewidth]{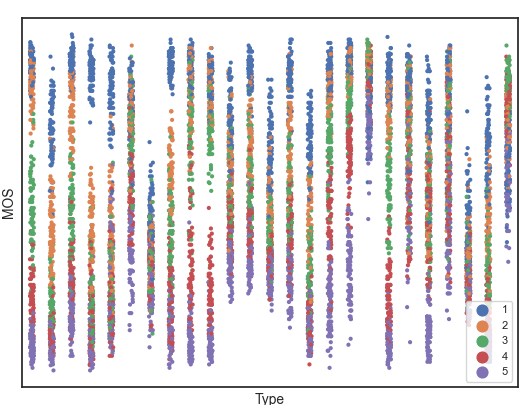}
    \label{subfig:ana-a}
    }
    \vspace{-0.3cm}
    \subfigure[]{
    \includegraphics[width=0.95\linewidth]{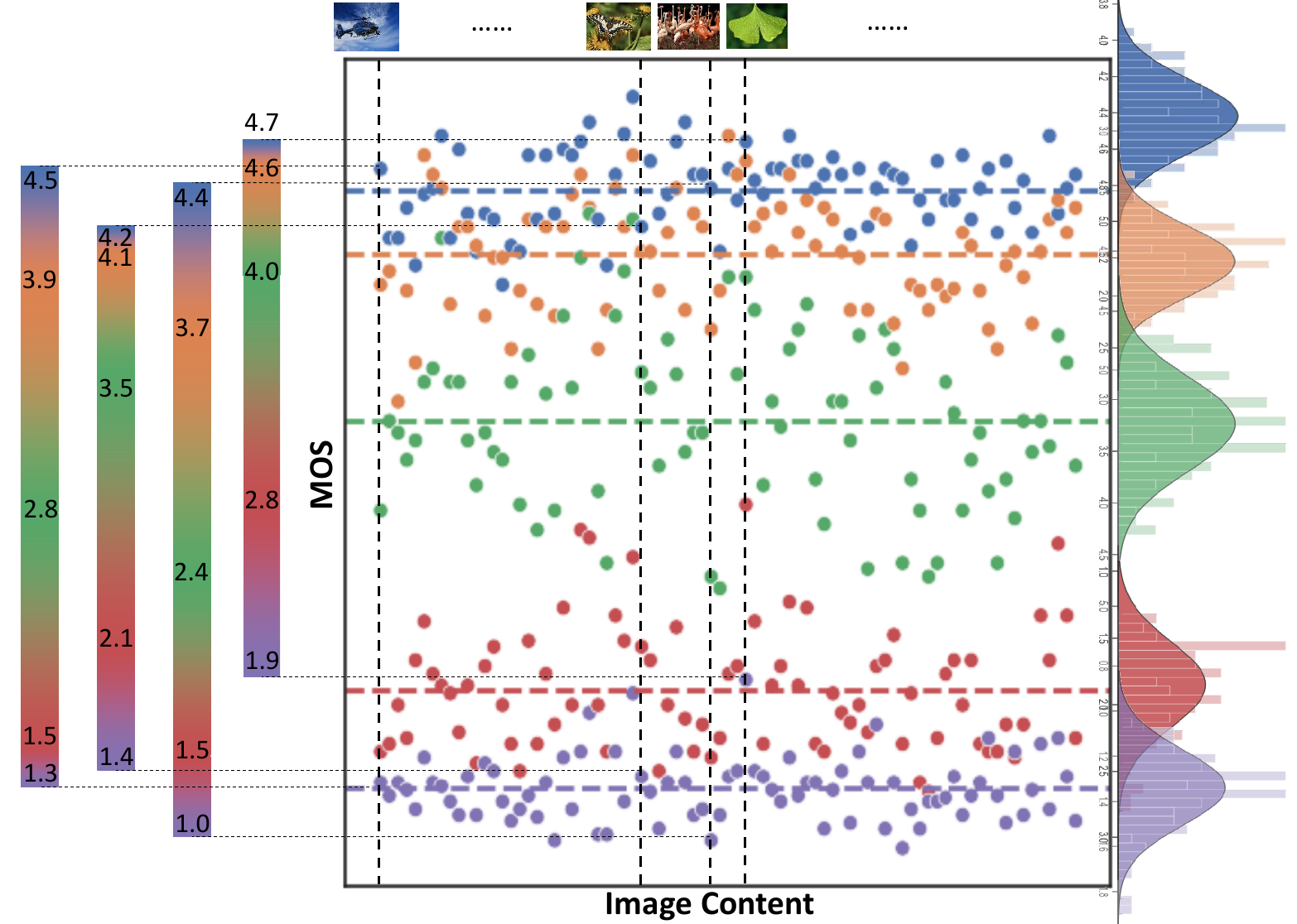}
    \label{subfig:ana-b}}
    \caption{Statistical analysis of Kadid-10k database. (a) shows the Type-MOS (Mean Opinion Score) distribution where different colors denote the distortion level.
    (b) the detailed distribution of a specific distortion type. 
    We select four image content as examples and illustrate their specific MOS under each level in the left, which shows that images with different content have different change in MOS when level changed from $1$ to $5$. We also show the MOS distribution of each level in the right, where each distribution generally tends to the Gaussian distribution.
}
\label{fig:analysis}

\end{figure}

\begin{figure*}[t]
\begin{center}
\includegraphics[width=0.9\linewidth]{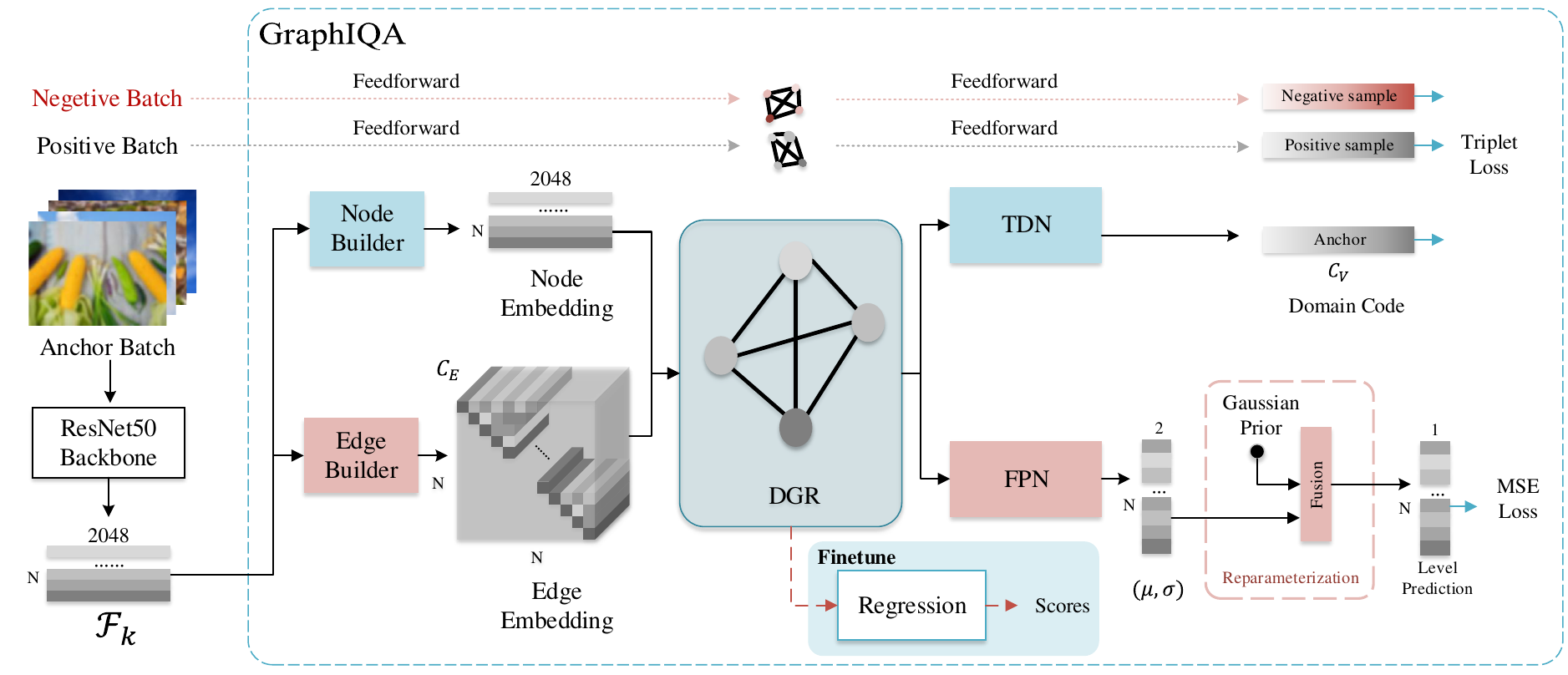}
\end{center}
   \caption{The illustration of proposed GraphIQA. We first train the networks to learn Distortion Graph Representations (DGRs). Each DGR is samples from one specific distortion type. For each iteration, three DGRs (Anchor batch, Positive batch and Negetive batch) are samples to train model to capture the relationship between distortion types with triplet loss. The learned DGRs are utilized to improve IQA performance by finetuning the regression network on a target IQA dataset. Note that GraphIQA doesn't require MOS or DMOS supervision in the first stage.}
\label{fig:framework}

\end{figure*}

\section{Methodology}
\label{method}

\subsection{Motivation}
\label{subsec:motivation}

Subjective image quality assessment is commonly obtained by collecting mean opinion scores from many subjects, which is labor-intensive and impractical. Recently, the learning-based methods have drawn much attention as the high efficiency and accuracy. As widely accepted, the human visual system has different sensitivity to different distortion types and levels~\cite{golestaneh2013no,hassen2013image}, and thus leveraging them to optimize the IQA task is a common approach. Most of the existing methods regard the different distortion types as a plane model, which is achieved by type or level classification task or learning to rank. 
They are proved to somehow bring in the improvements to IQA tasks, but fail to model the relationship between types and levels even other distortion related factors.

To further investigate how perceptual image quality is affected by distortion-related factors, we start from the analysis of IQA datasets. To get a more generalized conclusion, our analysis is based on Kadid-10k dataset~\cite{lin2019kadid}, which is a large scale dataset including $10,125$ images with $25$ distortion types and $5$ distortion levels. As observed from the statistics of Kadid-10k that is shown in Fig.~\ref{subfig:ana-a}, the distortion types are crucial influential factor to the distribution of IQA scores which is consistent with our common knowledge. Meanwhile, the distributions of diverse distortions also have difference, one of which is shown in Fig.~\ref{subfig:ana-b} in detail. As shown in Fig.~\ref{subfig:ana-b}, IQA scores present a sequential distribution according to different distortion levels, that is, the higher the level (means to be of more serious distortion) the lower the IQA scores. There is also a constant rule between samples with various levels, such as that the samples with level-$2$ is much more similar with samples with level-$1$ than them with level-$5$. In short, the distortion levels are the discrete points sampled on the degradation curve, in addition to their characteristics, they also satisfy the ranking relationship. Thus the level prediction problem is a regression problem.
Besides, the samples with the same distortion and level tend to have similar characteristic of distortion, so that the scores tend to cluster together.
This lead to the opinion that perceptual image quality is still affected by image content. According to our statistics of samples with same type and level, under the same type and level, the vibration of scores still exists, and it obeys the Gaussian distribution. 

In addition to the characteristics of each distortion-related factor, we also observe that the relationship between distortion type and distortion level tends to be a hierarchical model. That is, when we analyze the level, we tend to the hierarchical relationship where the IQA task first needs to consider type, and in each type, it needs further consider the level distribution of samples. This is more in line with the human eye’s analysis habit when analyzing distorted images, and is applied in some existing methods~\cite{liu2017rankiqa,xu2016multi}. Learning the ranking relationship between samples of totally different types and different levels will not provide prior knowledge for analysis, but will make it harder for analysis. Similar to the analysis of level, the prior knowledge of image content can be easier to capture when analyzing samples with fixed type and level.

Motivated by these observations, we conclude that the relationship between type and level is hierarchical and propose the use of the graph. Then we integrate the graph representation learning method to learn the distortion graph representations (DGRs). They can simultaneously represent the characteristic of each distortion and its internal structure related to the distribution of samples with different levels, so as to model the character of distortion-related factors and their hierarchical relationship at the same time. Correspondingly, we learn DGRs from two aspects, which are type discrimination task and fussy level prediction task. The detail will be described in detail in the following sub-sections.

\subsection{Distortion Graph Representation}
\label{sub:overview}

\begin{table}[]
\centering
\caption{The notations of important symbols.}
\begin{tabular}{c|l}
\hline
Symbol & Description                             \\ \hline
$G$      & Distortion representation graph (DGR) \\
$\mathcal{F}$ & The set of features obtained from backbone\\
$f$ & The specific feature obtained from backbone \\
$V$      & The set of Node embeddings in a graph \\
$v$      & The specific Node embedding             \\
$E$      & The union of Edge embeddings in a graph \\
$e$      & The specific Edge embedding             \\
$A$      & Adjacency matrix                        \\
$W_{EB}, W_{TDN}$      & Weights of GCN           \\
$y_{code}$ & The domain code \\
$y$ & The prediction of level \\ 
 \hline
\end{tabular}
\label{tab:notation}
\end{table}

We build DGR as shown in Fig.~\ref{fig:framework}, whose nodes represent samples, while edges indicate the relationships between each of them. The important symbols to be used in network together with their definitions are noted of in the TABLE~\ref{tab:notation}.
The DGR of distortion $k$ is formulated as $G_k=(V_k,E_k)$, in which the $V_k$ denotes the set of nodes and the $E_k$ denotes the set of edges to describe the relationship between nodes. Specifically, the anchor batch with $N$ samples from the same distortion type is first fed into a CNN backbone such as ResNet50~\cite{he2016deep} to obtain the feature set $\mathcal{F}_k=\{f_i|i=1,2,...,N\}$ where $f_i\in\mathbb{R}^{C}$ and $C$ is the feature dimension. The extracted feature $f_i$ from each sample is used as the initialization of the node in $V_k$, while the similarity between each node serves as the initialization of the edge in $\mathcal{E}_k$ which is commonly expressed as a 2D adjacency matrix $A_k\in\mathbb{R}^{N\times{N}}$. However, considering the complexity of the relationship between samples, we expand the 2D adjacency matrix to a 3D adjacency matrix $A_k\in\mathbb{R}^{N\times{N}\times{C_E}}$ where the representation of each edge is a vector with dimension size $C_E$ instead of a scalar. To build the DGRs, we design two learnable modules: Node Builder and Edge Builder respectively.

\paragraph{Node Builder (NB)} 
In DGRs, it is desirable that the representation of each node embedding should contain more distortion-related information so that it can be further used as an "clue" to distinguish from different distortion types. Therefore, we use a learnable network NB, composed of fully connected layers, to optimize node embedding, which is formulated as

\begin{equation}
    V_k=\{v_{k,i}|v_{k,i} = F_{NB}(f_{k,i};\theta), v_{k,i}\in\mathbb{R}^{C},i=1,2,...,N\},
\end{equation}
where $v_{k,i}$ denotes the node embedding of $i$-th sample, and $\theta$ denotes the network parameters of NB.


\paragraph{Edge Builder (EB)}
To obtain rich information about the contrast relationship between nodes, we expand the adjacency matrix to 3D, \ieno, we represent the connection between two nodes by a vector.
We take the  edge vectors $e^0_{k,i,j}\in\mathbb{R}^{C}$ as initial edge embedding $E_k^0$, which is the result of dot multiplication between each node embedding. 
The edge embedding is further optimized to represent internal structure by a graph convolution network (GCN)~\cite{kipf2016semi}. In detail, given the edge embedding $E_k^0$ and $A_{E_k}\in\mathbb{R}^{{N^2}\times{N^2}}$ as input, the process of computation of each layer for the GCN with $L$ layers can be formulated as:

\begin{equation}
    E_k^{l+1}=\text{ReLU}(\hat{A}_{E_k}{E_k^l}{W^l_{EB}}),
\end{equation}

where 

\begin{align}
     &D = \text{diag}(\sum\nolimits_{p=1}^{N^2}(A_{E_k}+I)_p),\\
    &\hat{A}_{E_k} = DA_{E_k}D.
    \label{eq:graph_norm}
\end{align}

The initialization of edge embedding $E_k^0$ serves as the input of the first layer of edge builder, and $E_k^l$ denotes the output of the GCN $l$-th layer. $W_l$ is the trainable parameter of GCN $l$-th layer. In the end, the optimized edge embedding of DGR is defined as:
\begin{equation}
    E_k =\{e_{k,i,j}|e_{k,i,j}\in\mathbb{R}^{C_E},i,j={1,2,...,N}\},
\end{equation}
where the $C_E$ is the dimension of edge embedding, which is set much smaller than $C$ to reduce computational complexity. 


\subsection{Domain Graph Optimization}
As shown in Fig.~\ref{fig:framework}, to equip the DGRs with the ability to both representing each distortion and the relationship between distortion levels, GraphIQA learns DGRs from the following two aspects. a) To learn the representation of distortion types that can be distinguished from other type, and the contrast relationship between them for better generalization, we design the TDN; b) To learn the distribution of distortion levels based on considering the content impact, we design the FPN.

\paragraph{Type Discrimination Network (TDN)}
\label{dist}

TDN is used to obtain the typical compact representation of each DGR, which helps to distinguish it from the others. Specifically, we design a GCN to aggregate global information from node embedding and relationships from edge embedding. The process is formulated as follow:
 
\begin{equation}
    V_k^{l+1} = \text{ReLU}(\hat{A}_{V_k}{V_k^l}{W^l_{TDN}}),
\end{equation}
in which the $V_k$ is the node embedding and the $A_{V_k}\in\mathbb{R}^{N\times{N}}$ is the adjacency matrix of nodes, which is calculated by transforming edge embedding $E_k$ through the average pooling across channels.The process of transforming edge embedding is noticed as Node Pooling. $\hat{A}_{V_k}$ is defined similar to Equation (\ref{eq:graph_norm}). The output of the TDN will be a vector with dimension $C_V$, named as code $y_{code}$. 
Then triplet loss $\mathcal{L}_{dist}$~\cite{schroff2015facenet} is utilized to learn the contrast representation of different distortion types. It is achieved by aggregating the anchor DGR and the DGR of the same distortion while separating it from the DGR of the other distortion. In detail, the forward propagation will be conducted three times to get triplet with three different input batches. The three sub-graphs are obtained by “Anchor Batch”, “Positive Batch ” and “Negative Batch” as it is shown in Fig.~\ref{fig:framework}. To simplify the diagram, we use dotted lines to represent the repeated forward propagation process. The loss function is formulated as:
\begin{equation}
\begin{split}
    \mathcal{L}_{dist} = max&(d(y_{code}^{Anchor},y_{code}^{+})\\
    &-d(y_{code}^{Anchor},y_{code}^{-})+ margin), 0),
\end{split}
\label{eq:tri}
\end{equation}
where $d$ denotes L2 distance, and $y_{code}^{+}$ is the DGR representing the same type with anchor DGR (positive sample in Fig.~\ref{fig:framework}) while $y_{code}^{-}$ is the DGR representing the different types (negative sample in Fig.~\ref{fig:framework}). 
Here, triplet loss can not only learn more subtle difference between distortion types, but also learn a contrast relationship between distortion types, which avoids the network overfitting to the distortion types in the training set.

\paragraph{Fuzzy Predictor Network (FPN)}
\label{sub:level}

We design an FPN to predict levels while considering the uncertainty caused by image content. Even if samples share the same distortion type and level, their perceptual image quality still is different due to their different content. According to Fig.~\ref{subfig:ana-b}, we assume that the scores of samples at the same level are distributed near the average score, obey the Gaussian distribution.
Then we perform the prediction by randomly sampling from the Gaussian prior distribution $\mathcal{N}(\mu, \sigma^2)$. As this process is not differentiable, the reparametrization trick~\cite{kingma2013auto} is used to ensure end-to-end training of the network. In detail, the $\epsilon$ is sampled from Normal distribution $\mathcal{N}(0,1)$, and then mapped to an arbitrary Gaussian distribution according to the generated hyper-parameters:

\begin{equation}
    y_{i} = \mu_{i} + \sigma_{i}\epsilon, \epsilon\in\mathcal{N}(0,1),
\end{equation}
where $\mu$ and $\sigma$ are the predicted mean and scale generated by the hyper predictor, as is shown in Fig.~\ref{fig:framework}. Because the prediction of distortion level is not only achieved by analyzing the distortion-related representation of nodes, but also the comparison between nodes to estimate the level of distortion, both the node embedding and edge embedding are needed. Specifically, node embedding $V_k$ is fed into FPN directly, while edge embedding $E_k$ is averaged across the rows, which is noticed as Edge Pooling, to aggregate the information from the current node and all the other neighboring nodes, as $E_k'=[\sum_{j}e_{k,i,j}]/N$. Then mean square error (MSE) loss function is utilized when training the hyper predictor:

\begin{equation}
    \mathcal{L}_{level} = \sum\nolimits_{i}|y_{i} - y'_{i}|^2,
\end{equation}
where $y'_{i}$ denotes the target level. The entire model will be trained end-to-end to minimize the combination of above loss functions, which are weighted by hyper-parameter $\lambda$:

\begin{equation}
    \mathcal{L} = \mathcal{L}_{dist} + \lambda\mathcal{L}_{level}.
\end{equation}

\begin{figure*}[h]
\centering
\subfigure[GraphIQA framework]{
\includegraphics[width=0.55\linewidth]{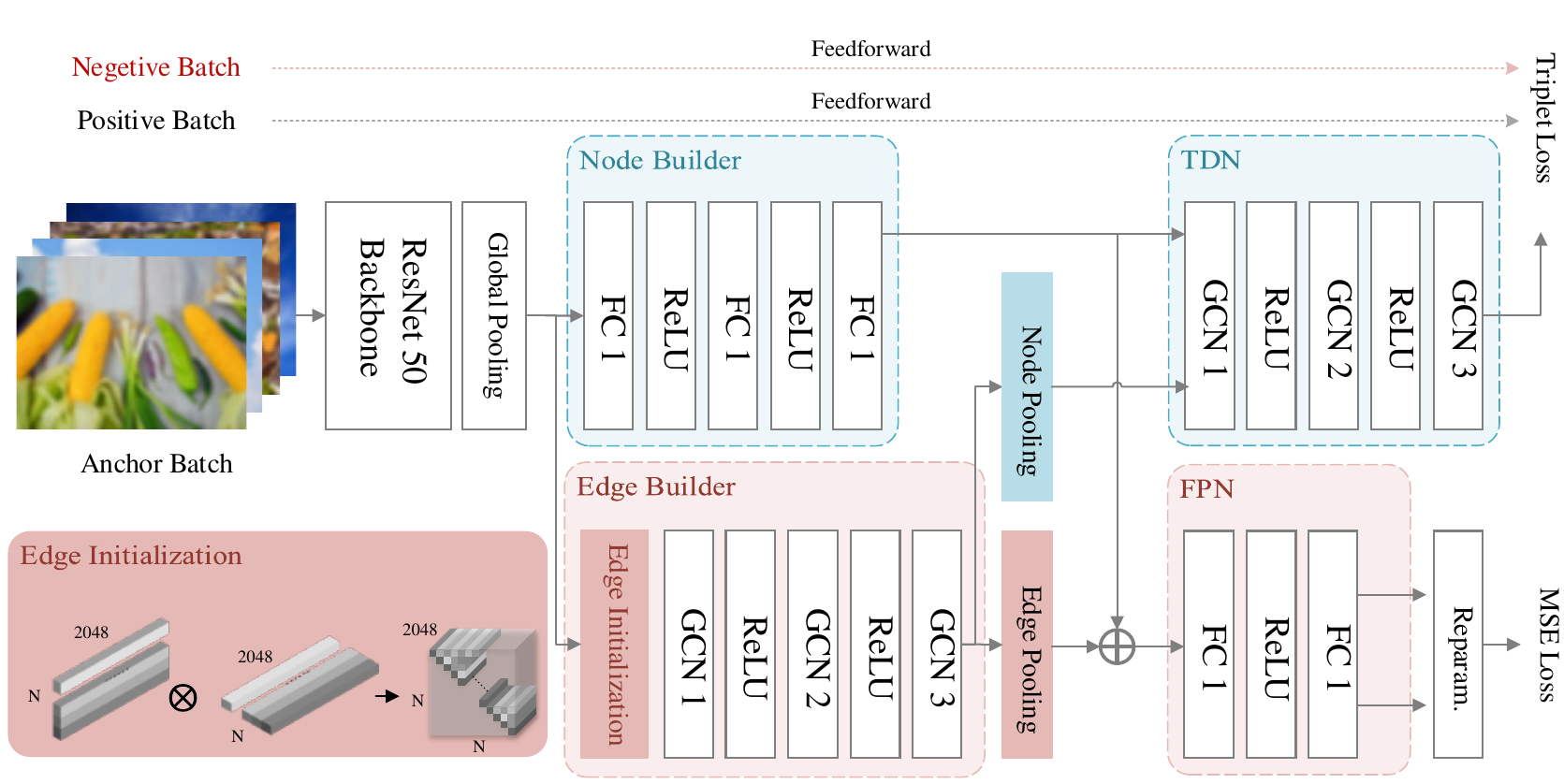}
\label{subfig:net-pretrain}}
\subfigure[Finetune GraphIQA on target database]{
\includegraphics[width=0.35\linewidth]{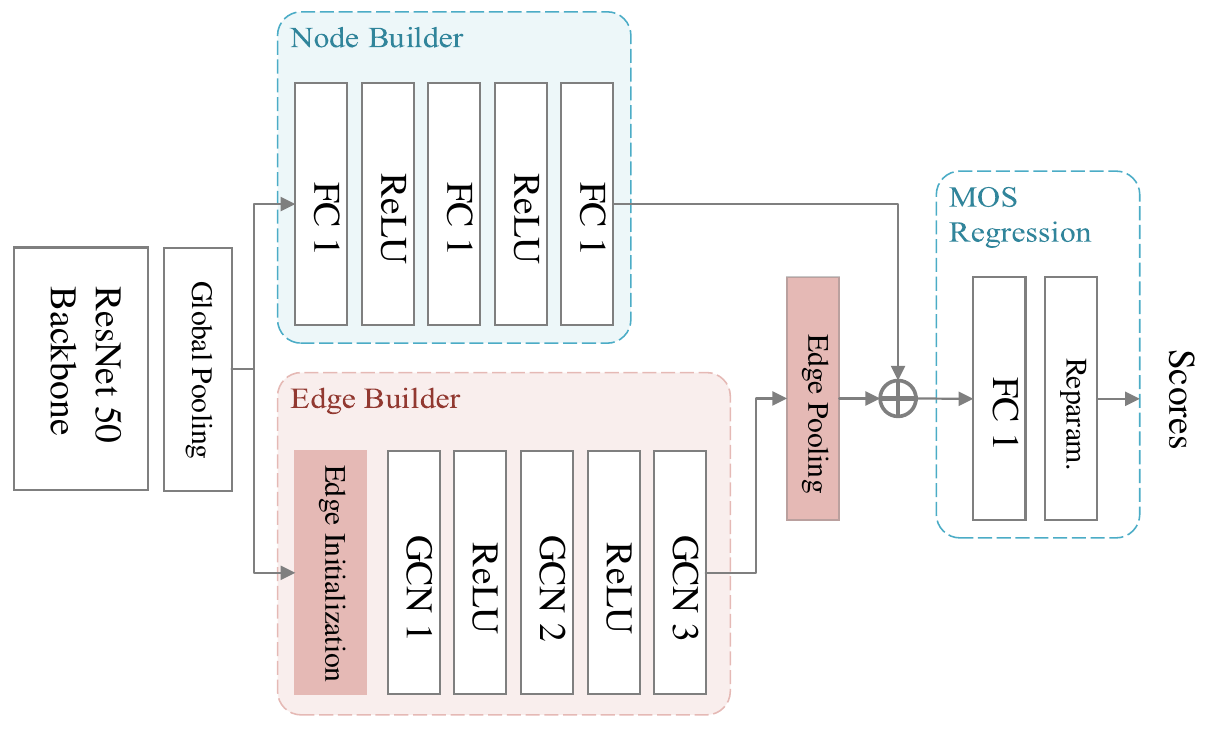}
\label{subfig:net-finetune}}
\caption{The illustration of the architecture of GraphIQA, in which (a) is the architecture for pre-training while (b) is the architecture for finetuning.}
\label{fig:network}
\end{figure*}

The detail of the architecture is shown in Fig.~\ref{fig:network}.

\subsection{Finetune and Inference}
\label{sub:inference}
Benefiting from the improved representation ability of DGRs, GraphIQA shows the potential for better fulfillment of IQA tasks. Specifically, when finetuning on the target dataset, both the node embedding and edge embedding are used to regress the IQA scores. And there is no need for representation building modules (\textit{i.e.}, TDN and FPN). Specifically, the node embedding and edge embedding is concatenated as a vector, and fed into the regression module. Here the edge embedding is self-loop edges, which contains rich distortion level priors learned from the pre-training phase.
The regression module is a small and simple network with two fully connected layers. When using DGRs for IQA on authentically distorted datasets, the prediction of authentic distortion datasets is achieved based on the Gaussian prior distribution so that it can better handle the unknown distortion type. Then the entire model is finetuned to minimize the MSE between ground truth (which is MOS/DMOS) and predicted scores, which is defined as:

\begin{equation}
    \mathcal{L}_{scores}=\frac{1}{N_f}\sum_{i=1}^{N_f}|c_i - c'_i|^2,
\end{equation}
where the $N_f$ denotes the mini-batch size. It is worth noting that as it already has the ability to infer the DGR, GraphIQA can support any size of input batch in the inference stage.



\begin{figure*}[h]
    \centering
    \includegraphics[width=1\linewidth]{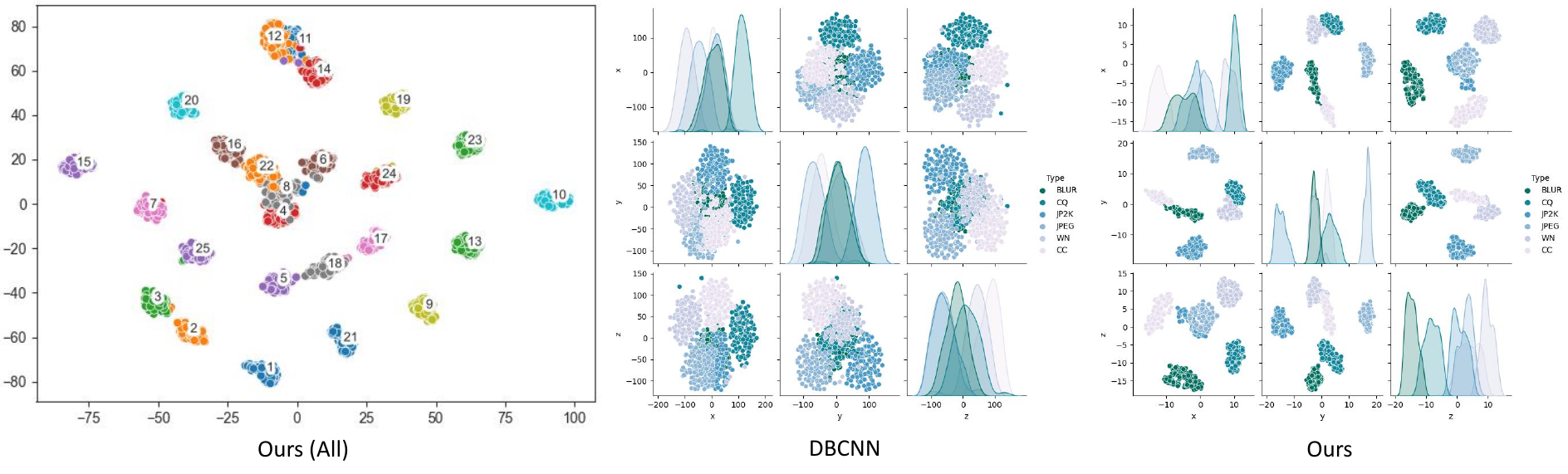}
    \caption{Visualization results of the representations which are generated by our method. To obtain the comparison results with DBCNN~\cite{zhang2018blind}, the visualization results of the representations of the common distortion types are shown in right sub-figure. It shows that the DGRs in our method are more discriminable than the representation learned by classification task in DBCNN.}
    \label{fig:visual}
\end{figure*}


\begin{table}[]
\centering
\caption{Evaluation on clustering performance of samples in each DGR on Kadid-10K dataset, which intuitively shows the performance of the DGR on level representation. The performance is measured by three metrics: homogeneity (means all of
the observations with the same class label are in the same
cluster, $1$ means the best), completeness (means all members of the same class
are in the same cluster, $1$ means the best) and V-measure (the combination of
both homogeneity and completeness, $1$ means the best).}
\label{tab:clustering}
 \setlength{\tabcolsep}{0.4mm}{
\begin{tabular}{c|ccc|c|ccc}
\toprule[1.5pt]
Type & Homo. & Comp. & V-m.  & Type     & Homo. & Comp. & V-m.  \\ \hline
GB   & 0.815 & 0.735 & 0.773 & MN       & 0.689 & 0.639 & 0.663 \\ \hline
LB   & 0.850 & 0.772 & 0.809 & Denoise  & 0.803 & 0.797 & 0.800 \\ \hline
MB   & 0.664 & 0.765 & 0.711 & Brighten & 0.573 & 0.511 & 0.524 \\ \hline
CD   & 0.424 & 0.449 & 0.437 & Darken   & 0.401 & 0.408 & 0.404 \\ \hline
CS   & 0.271 & 0.350 & 0.306 & MS       & 0.213 & 0.265 & 0.236 \\ \hline
CQ   & 0.557 & 0.583 & 0.570 & Jitter   & 0.706 & 0.635 & 0.669 \\ \hline
CSA1 & 0.554 & 0.571 & 0.562 & NEP      & 0.193 & 0.216 & 0.204 \\ \hline
CSA2 & 0.353 & 0.373 & 0.362 & Pixelate & 0.778 & 0.709 & 0.742 \\ \hline
JP2K & 0.549 & 0.614 & 0.580 & Quan.    & 0.373 & 0.385 & 0.379 \\ \hline
JPEG & 0.717 & 0.659 & 0.687 & CB       & 0.177 & 0.329 & 0.230 \\ \hline
WN   & 0.780 & 0.706 & 0.741 & HS       & 0.519 & 0.534 & 0.526 \\ \hline
WNCC & 0.662 & 0.794 & 0.772 & CC       & 0.300 & 0.427 & 0.452 \\ \hline
IN   & 0.772 & 0.712 & 0.741 &          &       &       &       \\ \bottomrule[1.5pt]
\end{tabular}}

\end{table}

\section{Experiments}
\label{exp}

\subsection{Experiments Setting}
\label{sub:exmp}

\paragraph{Dataset}
During pre-training, we use Kadid-10k or Kadis-700k~\cite{lin2019kadid}. The former is a large synthetic distorted database containing $81$ images with $25$ distortion types\footnote{GB: Gaussian blur; LB: Lens blur; MB: Motion blur; CD: Color diffusion; CS: Color shift; CQ: Color quantization; CSA1: Color saturation 1; CSA2: Color saturation 2; WN: White noise; WNCC: White noise in color component; IN: Impulse noise; MN: Multiplicative noise; MS: Mean shift; NEP: Non-eccentricity patch; Quan.: Quantization; CB: Color block; HS: High sharpen; CC: Contrast change} and $5$ distortion levels. The latter is a large-scale synthetic distortion dataset with $700,000$ distorted images with $25$ distortion types and $5$ distortion levels for each type. We use Kadid-10k dataset as validation set to select the hyper-parameters for pre-trained model.

For target datasets, we choose two datasets with authentic distortion (KonIQ-10k~\cite{lin2018koniq} and LIVE Challenge (LIVEC)~\cite{ghadiyaram2015massive}), two datasets with synthetic distortion (LIVE~\cite{sheikh2006statistical} and CSIQ~\cite{larson2010most}) and a dataset with multiple distortions (LIVEMD~\cite{jayaraman2012objective}). KonIQ-10k consists of $10073$ images which are selected from the large public multimedia database YFCC100m~\cite{thomee2016yfcc100m}. Those samples try to cover a wide and uniform quality distortion. LIVEC contains $1162$ images taken from different photographers with various cameras. LIVE contains $779$ images with $5$ distortion types and CSIQ contains $866$ images with $6$ distortion types. LIVEMD contains $450$ distorted images with $2$ multiple distortion types (i.e., blur-jpeg and blur-noise).
When finetuning, we split the dataset into a training set and a test set according to~\cite{zhang2018blind,su2020blindly}. Specifically, for synthetic distorted datasets, to avoid content overlapping between the training set and test set, we first randomly split the source images according to the ratio of $8:2$, and then assign the corresponding distorted images to obtain the training set and test set. For authentic distorted datasets, there is no image content overlapping, we directly split the whole dataset into a training set and a test set according to the ratio of $8:2$. All the results are obtained by training and test on the specific target dataset $10$ times with a randomly splitting operation, and the average results are reported.

\paragraph{Evaluation Metrics}
\label{sub:metric}

We mainly adopt two commonly used metrics, which are Spearman's rank order correlation coefficient (SRCC) and Pearson's linear correlation coefficient (PLCC) to measure the prediction monotonicity and prediction accuracy. Both of them range from $-1$ to $1$ and a higher value indicates better performance.

\paragraph{Implementation Details}
\label{par:arch}
We implement our model by PyTorch, and both training and testing are conducted on the NVIDIA 2080Ti GPUs. For data augmentation, when pre-training the GraphIQA model, we randomly sample from each distortion type and randomly crop them into $224\times224$ patches for 25 times, as there tends to be local distortion in the training database. The hyper-parameter $\lambda$ for loss function is set as $0.25$. The margin of the triplet loss function is set to $0.1$. We use Adam~\cite{kingma2014adam} optimizer to pre-train our representation model for $350000$ steps with mini-batch size of $32$. Learning rate is set to $1\times10^{-5}$. The dimension size of node embedding $C_V$ is set to $256$, and the size of edge embedding $C_E$ is set to $64$ (The detailed experiments on hyper-parameters are shown in Section~\ref{sub:arch}).
During finetuning, the input samples are randomly cropped into $224\times224$ at 10 times (some large images are resized to proper size firstly and randomly cropped to $224\times224$). We use Adam~\cite{kingma2014adam} optimizer to finetune on IQA task for $20$ epochs with the mini-batch size of $32$. The learning rate for finetuning is set to $5\times10^{-6}$. During the testing stage, all the testing images are randomly cropped to $10$ $224\times224$ patches, and their corresponding prediction scores are averaged to get the final quality scores. 

\begin{table*}[t]
\centering
\caption{SRCC comparison in cross distortion type on Kadid-10k dataset. We test
the IQA performance for one distortion type at a time, and performance for all the distortion types are shown in table. All the best results are highlighted in bold. }
\label{tab:kadid}
\begin{threeparttable}
\setlength{\tabcolsep}{1.1mm}{
\begin{tabular}{c|cccccccccccccc}
\toprule[1.5pt]
Dist. type & GB             & LB             & MB             & CD              & CS             & CQ             & CSA1           & CSA2           & JP2K           & JPEG           & WN             & \multicolumn{2}{c}{WNCC}           \\ \hline
WaDIQaM~\cite{bosse2017deep}    & 0.879          & 0.730          & 0.730          & 0.833           & 0.421          & 0.806          & 0.148          & 0.836          & 0.539          & 0.530          & 0.897          & \multicolumn{2}{c}{0.925}          \\ 
MetaIQA~\cite{zhu2020metaiqa}    & 0.946          & 0.917          & 0.926          & 0.892           & \textbf{0.785} & 0.717          & 0.304          & \textbf{0.931} & \textbf{0.945} & 0.912          & 0.905          & \multicolumn{2}{c}{0.930}          \\ 
Ours (w/o)* & 0.925          & 0.875          & 0.915          & 0.811           & 0.725          & 0.642          & \textbf{0.501} & 0.618          & 0.941          & 0.822          & 0.817          & \multicolumn{2}{c}{0.875}          \\ 
Ours       & \textbf{0.958} & \textbf{0.938} & \textbf{0.951} & \textbf{0.926}  & 0.738          & \textbf{0.873} & 0.462          & 0.929          & 0.938          & \textbf{0.944} & \textbf{0.916} & \multicolumn{2}{c}{\textbf{0.955}} \\ \midrule[1pt]
Dist. type & IN             & MN             & Denoise        & Brighten        & Darken         & MS             & Jitter         & NEP            & Pixelate       & Quan.   & CB             & HS               & CC              \\ \hline
WaDIQaM~\cite{bosse2017deep}    & 0.814          & 0.884          & 0.765          & 0.685           & 0.272          & 0.348          & 0.778          & 0.348          & 0.700          & 0.735          & 0.160          & 0.558            & 0.421           \\ 
MetaIQA~\cite{zhu2020metaiqa}    & \textbf{0.867} & 0.925          & 0.899          & 0.783           & 0.622          & 0.556          & 0.928 & 0.418 & 0.809 & \textbf{0.877} & 0.513 & 0.437            & 0.438           \\ 
Ours (w/o)* & 0.811          & 0.911          & 0.858          & 0.412 & 0.707 &0.071 & \textbf{0.949} & 0.541          & 0.800          & 0.639          & 0.334          & 0.749            & 0.049           \\ 
Ours       & 0.845 & \textbf{0.951} & \textbf{0.922} & \textbf{0.889}  & \textbf{0.806} & \textbf{0.745} & 0.943 & \textbf{0.677} & \textbf{0.873} & 0.867 & \textbf{0.626} & \textbf{0.904}   & \textbf{0.825}  \\  \bottomrule[1.5pt]
\end{tabular}}

\begin{tablenotes}
    \footnotesize
    \item * denotes the performance of the proposed GraphIQA without being finetuned on target datasets.  
\end{tablenotes}
\end{threeparttable}

\end{table*}

\begin{figure}[h]
    \centering
    \includegraphics[width=1\linewidth]{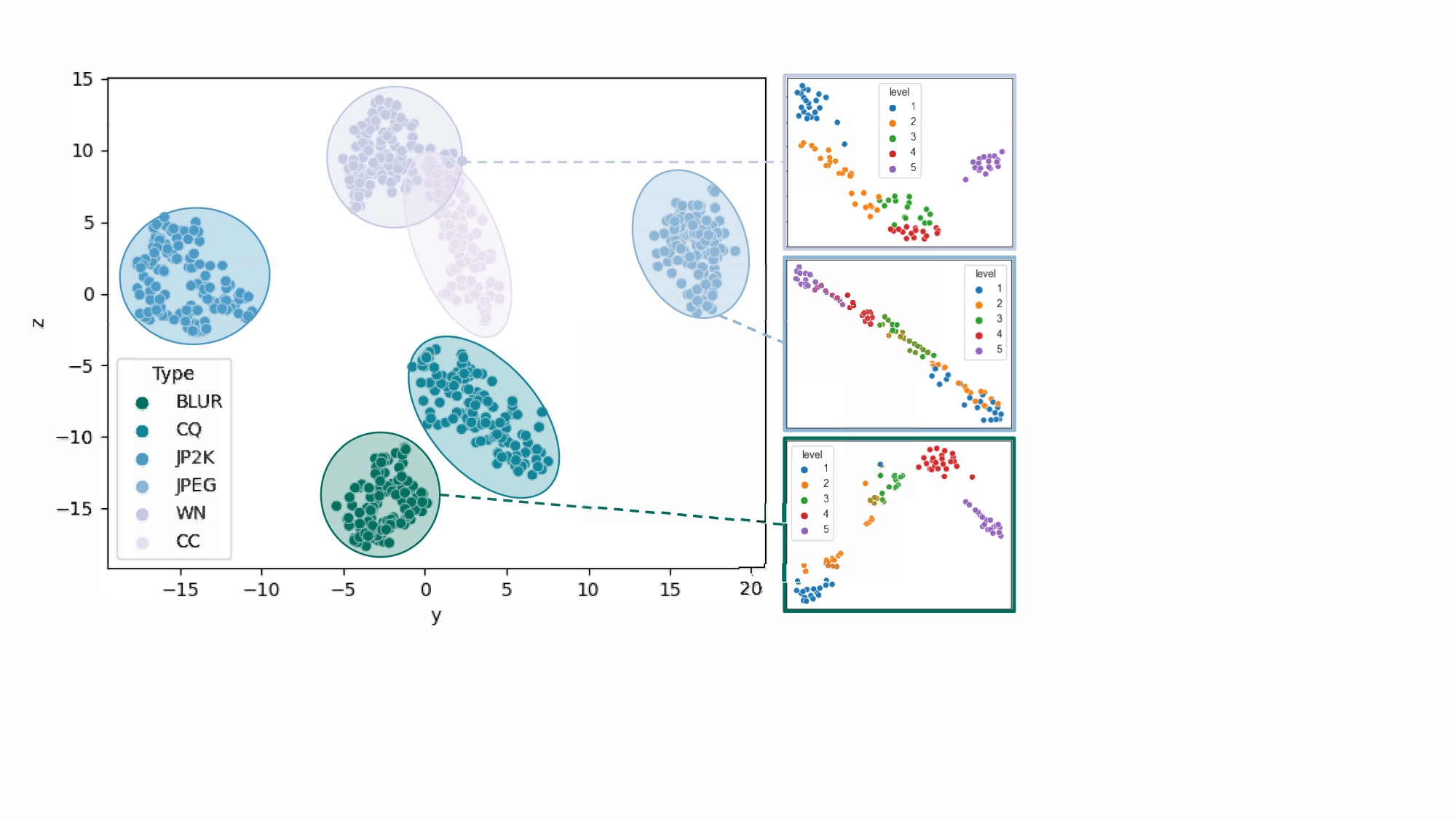}
    \caption{Visualization results of DGRs and its intrinsic structure. The DGRs themselves are highly discriminable. Besides, their internal distribution in accordance with the ranking relationship of distortion levels, indicating that the nodes of the DGR also have the ability to represent levels.}
    \label{fig:visual type and level}
\end{figure}

\begin{figure}[h]
    \centering
    \includegraphics[width=1\linewidth]{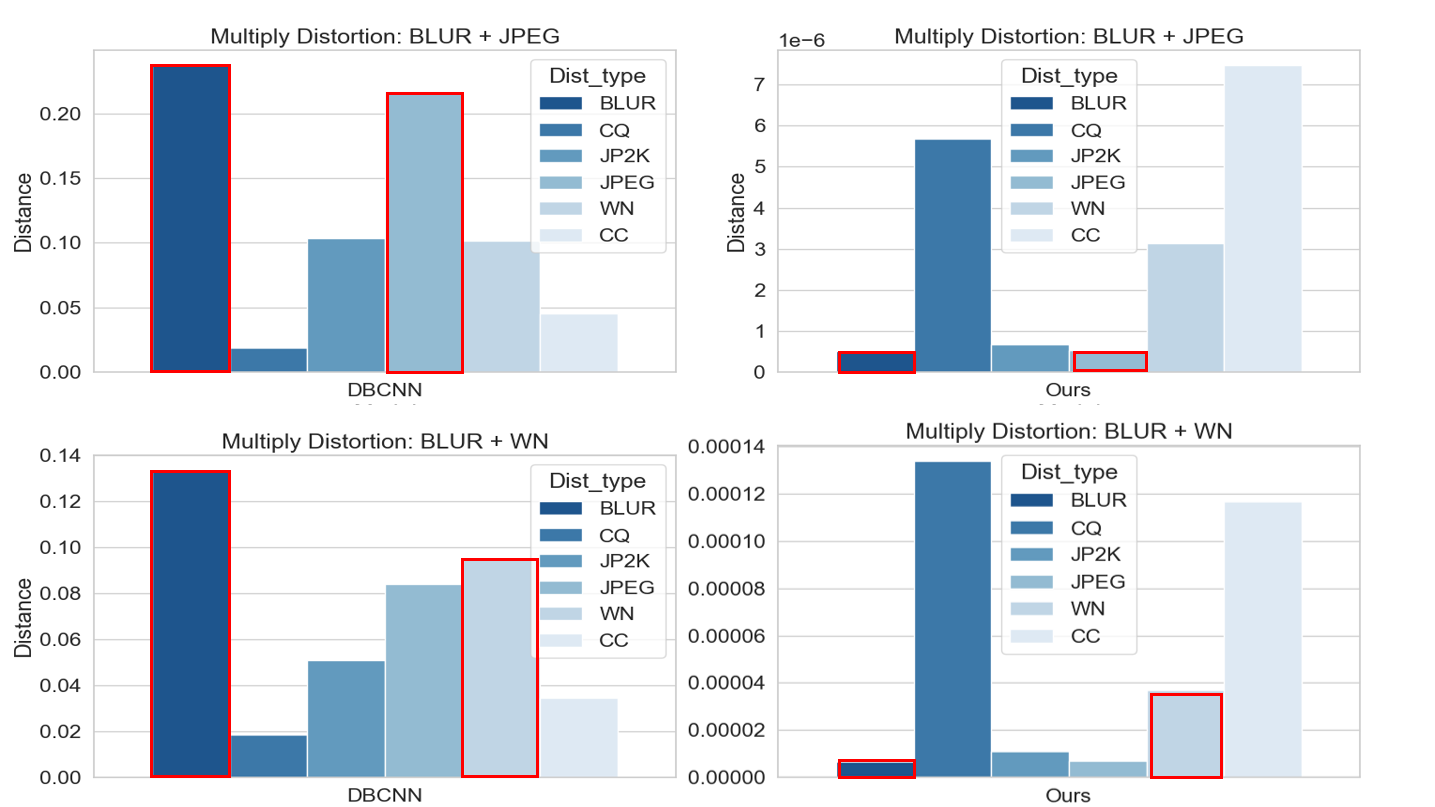}
    \caption{Visualization of interpretability for unseen distortion. This figure shows the distance between representations of the unseen multiple distortion type, \ieno, blur and jpeg (noted as BLUR+JPEG), and blur and noise (BLUR+WN). It proves that our method has the potential of interpretability that the representations of BLUR+JPEG are closest to them of BLUR and JPEG. and the representations of BLUR+WN are closest to them of BLUR.}
    \label{fig:visual distance}
\end{figure}

\subsection{DGR Performance Evaluation}
\label{sub:performance of DGR}
We evaluate the effectiveness of the proposed DGR from three aspects. 

\paragraph{Visualization and Clustering Evaluation} 
We visualize distribution of learned DGRs using \textit{t-SNE}~\cite{linderman2019fast}, as are shown in Fig.~\ref{fig:visual} and ~\ref{fig:visual type and level} respectively, which is then compared with representations generated by DBCNN~\cite{zhang2018blind}. We randomly sample $128$ times from six typical distortion types in Kadid-10k dataset and feed into Kadis-700k pre-trained model (without image content overlap) to get the dimension-reduced embedding. To reduce the bias of the visualization caused by \textit{t-SNE} during dimension reducing, we visualize the distribution of three-dimension space.
We can see that the DGRs are well-clustering representation according to their corresponding distortion types on the whole, but representations generated by DBCNN are less discriminable.
We also visualize the internal distribution of each DGR, as shown in Fig.~\ref{fig:visual type and level}. From them, we can observe that the node embeddings are not only clustering well according to the distortion level, but also show a regular pattern according to the order of levels. TABLE~\ref{tab:clustering} provides the clustering performance for all $25$ distortion types in Kadid-10k dataset, which is measured by homogeneity, completeness and V-measure. All of the metrics used for measurement of clustering performance is ranged from $0$ to $1$, and the $0$ means poor clustering performance while the $1$ mean the best. 
Considering that the prediction of levels is sampled from Gaussian prior distribution to model the influence by image content, the high accuracy is not our main concern. 
In TABLE~\ref{tab:clustering}, most of the results are higher than $60\%$, which further proves the powerful representation capability of each DGR to characterize each level.

\paragraph{Interpretability for unseen distortions}
To verify that the DGRs have better generalization capability, we compare the interpretability for unseen distortions of our models and DBCNN. We test on multiple distortion dataset (LIVEMD). There are two multiple distortion types which are BLUR and JPEG, and BLUR and white noise.We first test the interpretability only on the former multiple distortion type. The distance between embedding of samples of multiple distortions and every single distortion is calculated, which is shown in Fig.~\ref{fig:visual distance}. With our method, the distance between BLUR+JPEG and BLUR is the closest, and JPEG is the second closest. However, the representation generated by DBCNN is not interpretable. For the distortion BLUR+WN,  though considering that features of the latter are highly unstable with different mixing method (e.g., mixing order), it also shows the potential of interpretability. That is, when compared to the distance between the BLUR+JPEG distortion type and Noise, the distance between the BLUR+WN distortion type and Noise is much closer.
It shows that benefit from the learning of contrastive relationships between distortion types, our method is able to learn a more robust representation that even can somehow handle the unseen distortion types.

\paragraph{Leave-One Evaluation on Kadid-10k Dataset}
To further validate the contribution of DGRs to IQA task, we test the IQA performance for each distortion type. We compare our method with two CNN based BIQA methods by using the Leave-One-Distortion-Out cross validation. In detail, there is one distortion type left for testing and the rest of types are used as training set. All of the results are obtained by using the source code provided by their authors in the same training-testing conditions. With all the best results highlighted in bold, we can see from the TABLE~\ref{tab:kadid} that GraphIQA scheme can already achieve competing performance without finetuning on the target dataset. After finetuning on the training set, the performance can get even better that we reach the best on most of the distortion types ($18$ out of $25$).

\subsection{Comparison with the State-of-the-arts}
\label{sub:compare}
\begin{table}[t]
\centering
\caption{Comparison with SOTA methods on multiple datasets with SRCC and PLCC metrics. Among the compared methods we focus on comparing with the classical ones of using representational learning to improve IQA performance. These methods are listed at the bottom of the tables.}
\label{tab:compare}
\begin{tabular}{c|lllll}
\toprule[1.5pt]
\multicolumn{1}{l|}{}         & \multicolumn{5}{c}{SRCC}           \\ \hline
                              & KonIQ & LIVEC & LIVE & CSIQ & LIVEMD   \\ \hline
BRISQUE~\cite{mittal2012no}               & 0.665 & 0.608 & 0.939 & 0.746 & 0.886 \\
ILNIQE~\cite{zhang2015feature}& 0.507 & 0.432 & 0.902 & 0.806 & 0.876 \\
HOSA~\cite{xu2016blind}       & 0.671 & 0.640 & 0.946 & 0.741 & 0.913  \\
BIECON~\cite{kim2016fully}                & 0.618 & 0.595 & 0.961 & 0.815 & 0.909   \\
WaDIQaM~\cite{bosse2017deep}  & 0.797 & 0.671 & 0.954 & \textbf{0.955} & -   \\
SFA~\cite{li2018has}                   & 0.856 & 0.812 & 0.883 & 0.796 & -  \\
PQR~\cite{zeng2017probabilistic}  & 0.880 & \textbf{0.857} & 0.965 & 0.873 & -  \\
UNIQUE~\cite{zhang2021uncertainty} & 0.896 & 0.854 & 0.969 &0.902 & - \\
HyperIQA~\cite{su2020blindly} & 0.905 & \textbf{0.856} & 0.962 & 0.920 & - \\
MetaIQA~\cite{zhu2020metaiqa}  &  0.850 & 0.802 & - & - & - \\
MetaIQA+~\cite{zhu2021generalizable}  &  \textbf{0.909} & 0.852 & - & - & - \\
CaHDC~\cite{wu2020end} & - & - & 0.965 & 0.903 & \textbf{0.927} \\

\hline
CNNIQA++~\cite{kang2015simultaneous} & - & - & 0.965 & 0.892 & \textbf{0.927}  \\
RankIQA~\cite{liu2017rankiqa} &      -  &   0.641  &  \textbf{0.981} &0.892& 0.908     \\
MEON~\cite{ma2018end} &  -  & - & 0.951 & 0.852 & 0.924  \\
DBCNN~\cite{zhang2018blind}   & 0.872 & 0.852 & 0.967 & 0.946 & \textbf{0.927}   \\

Ours     & \textbf{0.911}  & 0.845 & \textbf{0.979} & \textbf{0.947} & \textbf{0.930} \\\midrule[1pt]
\multicolumn{1}{l|}{} & \multicolumn{5}{c}{PLCC} \\ \hline
                     & KonIQ & LIVEC & LIVE & CSIQ &  LIVEMD\\ \hline
BRISQUE~\cite{mittal2012no} &  0.681 & 0.629 & 0.935 & 0.829 & 0.917\\
ILNIQE~\cite{zhang2015feature} &  0.523 & 0.508 & 0.865 & 0.808 & 0.863\\
HOSA~\cite{xu2016blind} &  0.694 & 0.678 & 0.947 & 0.823 & 0.926 \\
BIECON~\cite{kim2016fully} &  0.651 & 0.613 & 0.962 & 0.823 & 0.933\\
WaDIQaM~\cite{bosse2017deep} &  0.805 & 0.680 & 0.963 & \textbf{0.973}& - \\
SFA~\cite{li2018has} &  0.872 & 0.833 & 0.895 & 0.818 & - \\
PQR~\cite{zeng2017probabilistic}  &  0.884 & \textbf{0.882} & 0.971 & 0.901 & - \\
UNIQUE~\cite{zhang2021uncertainty} & 0.876 & 0.890 & 0.968 &0.927 & - \\

HyperIQA~\cite{su2020blindly} &  \textbf{0.922} & \textbf{0.882} & 0.966 & 0.943 & -  \\
MetaIQA~\cite{zhu2020metaiqa}  &  0.887 & 0.835 & - & - & - \\
MetaIQA+~\cite{zhu2021generalizable}  &  \textbf{0.922} & 0.852 & - & - & - \\
CaHDC~\cite{wu2020end} & - & - & 0.964 &0.914 & \textbf{0.950} \\

\hline
CNNIQA++~\cite{kang2015simultaneous} & - & - & 0.966 & 0.905 & 0.924   \\
RankIQA~\cite{liu2017rankiqa} & -  &  0.675 & \textbf{0.982}  & 0.912 & 0.929   \\
MEON~\cite{ma2018end} &  -  & - & 0.955  & 0.864 & \textbf{0.940} \\
DBCNN~\cite{zhang2018blind} &  0.881 & \textbf{0.865} & 0.971 & \textbf{0.959} & 0.934 \\
Ours &  \textbf{0.915} & 0.862 & \textbf{0.980} & \textbf{0.959} & \textbf{0.940} \\\bottomrule[1.5pt]
\end{tabular}

\end{table}

\begin{table*}[h]
\centering
\caption{SRCC comparison on individual type in LIVE and CSIQ dataset.}
\label{tab:individual}
\setlength{\tabcolsep}{1.5mm}{
\begin{tabular}{c|lllll|l|llllll|l}
\toprule[1.5pt]
\multicolumn{1}{c|}{Dataset}  & \multicolumn{6}{c|}{LIVE}  & \multicolumn{7}{c}{CSIQ}       \\ \hline
Type                          & JP2K  & JPEG  & WN    & GB    & FF    & Total & JP2K  & JPEG  & WN    & GB    & CC    & FN & Total\\ \hline
BRISQUE~\cite{mittal2012no}                      & 0.929 & 0.965 & \textbf{0.982} & 0.964 & 0.828 & 0.939 & 0.840 & 0.806 & 0.723 & 0.820 & 0.804 & 0.378 & 0.746 \\
ILNIQE~\cite{zhang2015feature}& 0.894 & 0.941 & 0.981 & 0.915 & 0.833 & 0.902 & 0.906 & 0.899 & 0.850 & 0.858 & 0.501 & 0.874 & 0.806 \\
HOSA~\cite{xu2016blind}       & 0.935 & 0.954 & 0.975 & 0.954 & 0.954 & 0.946 & 0.818 & 0.733 & 0.604 & 0.841 & 0.716 & 0.500 & 0.741 \\
BIECON~\cite{kim2016fully}                       & 0.952 & 0.974 & 0.980 & 0.956 & 0.923 & 0.961 & 0.954 & 0.942 & 0.902 & 0.946 & 0.523 & 0.884 & 0.815  \\
WaDIQaM~\cite{bosse2017deep}  & 0.942 & 0.953 & 0.982 & 0.938 & 0.923 & 0.954 & 0.947 & 0.853 & \textbf{0.974} & \textbf{0.979} & 0.923 & 0.882 & \textbf{0.955}   \\
HyperIQA~\cite{su2020blindly} & 0.949 & 0.961 & 0.982 & 0.926 & 0.936 & 0.962  & \textbf{0.960} & 0.934 & 0.927 & 0.915 & 0.874 & 0.931 & 0.920   \\ \hline
DBCNN~\cite{zhang2018blind}   & 0.955 & 0.972 & 0.980 & 0.935 & 0.930 & 0.967 & 0.953 & 0.940 & 0.948 & 0.947 & 0.870 & 0.940 & 0.946   \\
Ours (w/o)*                    & 0.901 & 0.825 & 0.385 & 0.791 & 0.908 & 0.711 & 0.859 & 0.912 & 0.883 & 0.800 & 0.029 & 0.854 & 0.705   \\
Ours                      & \textbf{0.979} & \textbf{0.978} & 0.978 & \textbf{0.978} & \textbf{0.979} & \textbf{0.979}  & 0.947 & \textbf{0.947} & 0.948 & 0.947 & \textbf{0.947} & \textbf{0.948} & 0.947   \\ \bottomrule[1.5pt]
\end{tabular}}
\begin{tablenotes}
    \footnotesize
    \item * denotes the performance of the proposed GraphIQA without being finetuned on target datasets.  
\end{tablenotes}

\end{table*}

We compare our GraphIQA with the state-of-the-art (SOTA) BIQA methods including hand-craft feature based methods~\cite{mittal2012no,zhang2015feature,xu2016blind}, deep learning based synthetic IQA methods~\cite{bosse2017deep,kim2016fully,kang2015simultaneous,ma2018end,liu2017rankiqa,wu2020end} and deep learning based authentic IQA methods~\cite{zeng2017probabilistic,li2018has,zhang2018blind,su2020blindly,zhu2021generalizable,zhang2021uncertainty,zhu2020metaiqa}. All of the experiments are conducted $10$ times to avoid the bias of randomness.

\paragraph{Single Database Evaluations}
The results are shown in TABLE~\ref{tab:compare}, and the best results are highlighted in bold. Our approach outperforms all of the methods on both synthetic distortion and multiple distortion datasets (LIVE, CSIQ and LIVEMD). And for authentic distortion datasets (KonIQ-10k and LIVEC), Our method can obtain comparable performance with methods that are well designed for authentic distorted data~\cite{su2020blindly}. Notably, in our ours method, none of the authentic distorted data and multiply distorted data is used in pre-training stage. This shows that our pre-trained model shows better generalization ability to be transferred to other distortion domain. Observed from the results with pre-trained models which are trained on Kadid-10k dataset and Kadis700k dataset, with a large-scale dataset the pre-trained model with proposed method can get better performance, especially on synthetic distorted data. 
This suggests that the learned DGRs can be utilized to deal with both synthetically and authentically distorted images. 

We also present the performance comparison of our approach on individual distortion types. We choose LIVE and CSIQ which are unseen in our pre-training stage for a fair comparison. The results are shown in TABLE~\ref{tab:individual}. Compared with some methods, our approach, which is pre-trained without using the annotation of MOS/DMOS, noticed as Ours (w/o) in table, can still get comparable performance on some distortion types. After finetuning on the target dataset MOS/DMOS annotations, the performance get better and more consistent over individual distortion types.
This shows that the DGRs do provide rich effective prior for IQA, while having better generalization.

\paragraph{Generalization Evaluation}
At first, we run the cross dataset tests on both synthetically distorted dataset pair (LIVE and CSIQ) and authentically distorted dataset pair (KoinIQ and LIVE). We select the most competing methods, DBCNN~\cite{zhang2019learning} and HyperIQA~\cite{su2020blindly} for comparison. In the implementation, we use one dataset as a training set and the other one is used as a testing set. The results are shown in TABLE~\ref{tab:cross}, it can be observed that our approach can get comparable performance with the other methods. This is because of the strong generalization ability of our approach.

Then, to explore the generalization of GraphIQA on the quality assessment of enhanced images, we compare the performance on de-hazed quality assessment dataset ~\cite{min2018objective} and de-raining quality assessment dataset~\cite{wu2020subjective}, which is shown in TABLE~\ref{tab:enhan}. As is shown, our method can achieve better performance on de-hazed dataset and comparable performance on de-raining dataset when compared with Res50 (the same backbone), though the enhanced image is un-known for pre-training and there is no design targeting the distortion properties on enhanced images in our method. This verifies the generalization of GraphIQA.


\begin{table}[]
\centering
\caption{Cross-dataset evaluation to verify the generalization.}
\label{tab:cross}

\begin{tabular}{l|l|ll|l}
\toprule[1.5pt]
\multicolumn{2}{c|}{Dataset} & \multicolumn{1}{c}{\multirow{2}{*}{DBCNN}} & \multicolumn{1}{c|}{\multirow{2}{*}{HyperIQA}} & \multicolumn{1}{c}{\multirow{2}{*}{Ours}} \\ \cline{1-2}
Train         & Test         & \multicolumn{1}{c}{}                       & \multicolumn{1}{c|}{}                          & \multicolumn{1}{c}{}                      \\ \hline
KonIQ         & LIVEC        & 0.734 & 0.773 & \textbf{0.798}                                    \\
LIVEC         & KonIQ        & 0.\textbf{788} & 0.733 & 0.773                                    \\ \hline
CSIQ          & LIVE         & 0.909 & 0.940 & \textbf{0.945}                                    \\
LIVE          & CSIQ         & 0.775 & \textbf{0.834} & 0.830                                    \\ \bottomrule[1.5pt]
\end{tabular}

\end{table}

\begin{table}[]
\centering
\caption{Evaluation on enhanced quality assessment dataset to verify the generalization}
\begin{tabular}{l|ll|ll}
\toprule[1.5pt]
 & \multicolumn{2}{c|}{\begin{tabular}[c]{@{}c@{}}De-Hazed\\ Dataset\end{tabular}} & \multicolumn{2}{c}{\begin{tabular}[c]{@{}c@{}}De-Raining\\ Dataset\end{tabular}} \\ \hline
             & SRCC   & PLCC   & SRCC   & PLCC   \\ \hline
VGG19        & 0.8258 & 0.8310 & 0.4145 & 0.4139 \\
Res50        & 0.8361 & 0.8422 & 0.4299 & 0.4231\\
Ours (Res50) & 0.8386 & 0.8497 & 0.4237 & 0.4179 \\ \bottomrule[1.5pt]
\end{tabular}
\label{tab:enhan}
\end{table}

\begin{table}[]
\centering
\caption{SRCC evaluation of ablation study on KonIQ and LIVE dataset. Res50 denotes the ResNet model. In pre-training setting, the marked part means it is pre-trained with distortion type and level on Kadid10k dataset. In finetuning setting, the marked part means it is used to predict final scores.}
\setlength{\tabcolsep}{1mm}{
\begin{tabular}{llllllll|ll}
\toprule[1.5pt]
\multicolumn{8}{c|}{Setting}                                                          & \multicolumn{2}{c}{\multirow{2}{*}{Dataset}} \\ \cline{1-8}
\multicolumn{4}{c|}{Pre-train}                 & \multicolumn{4}{c|}{Finetune}         & \multicolumn{2}{l}{}                         \\ \hline
Res50 & DGRs & FPN & \multicolumn{1}{l|}{cls} & Res50 & DGRs & Nodes & Edges & KonIQ                 & LIVE                 \\ \hline
      &      &     & \multicolumn{1}{l|}{}    & \checkmark     &      &            &           & \textbf{0.904}                 & 0.964                \\
\checkmark     & \checkmark     & \checkmark    & \multicolumn{1}{l|}{}    & \checkmark     &      &            &           & 0.899                 & 0.956                \\
\checkmark& \checkmark& \checkmark& \multicolumn{1}{l|}{}    & \checkmark     & \checkmark    & \checkmark          &           & 0.898                 & 0.970                \\
\checkmark     & \checkmark    & \checkmark   & \multicolumn{1}{l|}{}    & \checkmark    & \checkmark &            & \checkmark      & 0.298                 & 0.855                \\
\checkmark& \checkmark&     & \multicolumn{1}{l|}{\checkmark}   & \checkmark& \checkmark& \checkmark & \checkmark& 0.899                 & 0.969                \\
\checkmark& \checkmark& \checkmark& \multicolumn{1}{l|}{}    & \checkmark& \checkmark& \checkmark& \checkmark& \textbf{0.911}        & \textbf{0.979}       \\\bottomrule[1.5pt]
\end{tabular}}
\label{tab:ablation}
\end{table}



\begin{table}[h]
\centering
\caption{The linear evaluation of the dimension size of the edge and the node on Kadid-10k dataset.}
\begin{tabular}{l|lllll}
\toprule[1.5pt]
Edge & 1     & 16    & 32    & 64    & 96    \\ \hline
SRCC    & 0.8246 & \textbf{0.8316} & 0.7015 & 0.8270 & 0.8137 \\
\hline
Node & 32    & 64    & 128   & 256   & 300   \\ \hline
SRCC    & 0.8227 & 0.8034 & 0.7932 & \textbf{0.8270} & 0.8079 \\
\bottomrule[1.5pt]
\end{tabular}

\label{tab:edge}
\end{table}

\subsection{Ablation Study}

\label{sub:ablation}
To evaluate the efficiency of each component in our approach. We conduct ablation study on both synthetic and authentic distortion datasets, \ieno, KonIQ and LIVE. And when select the size of node embedding and edge embedding, we conduct the linear evaluation experiments on validation set, Kadid-10k dataset, as~\cite{he2020momentum}, where the parameters of DGRs generation part (Backbone, Node Builder and Edge Builder) are fixed and the linear convolutional layers are trained to regress the MOS. The SRCC results are reported.

\paragraph{Model Components} As the scheme of using backbone ResNet50 is treated as the baseline, all the components are integrated to it, as shown in TABLE ~\ref{tab:ablation}. We first verify the efficiency of each component in the pre-training stage. It is observed that ImageNet pre-trained model can get better performance on the authentically distorted images, but inferior performance on synthetically distorted images. However, training the distortion representation without the model of a graph may degrade the performance. When training the representation for level, as the prediction of level is a regression problem instead of a classification task, the performance of using FPN is better than using the softmax classifier.
Than we verify the effectiveness of DGR and its components. 
The performance of utilizing pre-trained backbone together with edge embedding and node embedding separately is not as good as using them in combination, especially for edge embedding. This is because the most of information on edges has been aggregated into nodes during the training stage. When combining both of them, the GraphIQA can get better performance compared with baseline.
Then we further provide the performance on ImageNet pre-trained backbone. As is shown in the table, the ImageNet pre-trained backbone performs better on authentically distorted data while worse on synthetically distorted data which is consistent with the comments in~\cite{zhang2018blind}. 
However, with our method, the pre-trained model can obtain better improvement on synthetic distortion dataset while maintaining the performance on authentic distortion dataset. What is worth noting is that only the single pre-trained model trained with synthetic data is used in our method. It is shown that with proposed method, the better representation of synthetically distorted data can be learned which is par for the course. Meanwhile, as the representations is obtained by learning the relative relationship between synthetic distortion types, the representation is robust to various distortion types than that obtained by classification task, so that it can be transferred to authentically distorted data much easier.

\paragraph{Edge Embedding Size and Node Embedding Size} Then, we compare the performance on different sizes of edge embedding in TABLE~\ref{tab:edge}, which are $1$, $16$, $32$, $64$ and $96$. We can see that DGRs with edge embedding set as $16$ can get better performance, which demonstrates the effectiveness of the proposed 3D adjacency matrix. As for node embedding size, we compared the performance on size $32$, $64$, $128$, $256$ and $300$. And the performance is the best when it is set as $256$. It is noted that when one hyper-parameter is under selection, the others are set as the initial hyper-parameter set, where node embedding size is set as $256$, edge embedding size is set as $64$ and margin size is set as $0.2$.


\subsection{Experiments on architecture and hyper-parameters.}
\label{sub:arch}

\begin{table}[h]
\centering
\caption{Exploring the best architecture of our model.}
\begin{tabular}{c|cccc}
\toprule[1.5pt]
\multicolumn{5}{c}{Node Builder}  \\ \hline
FC layer  & 1      & 2      & 3      & 4      \\ \hline
SRCC     & 0.8021    & 0.8052    & \textbf{0.8270} & 0.7988 \\
\hline
\hline
\multicolumn{5}{c}{Edge Builder}              \\ \hline
GCN layer & 1      & 2      & 3      & 4      \\
\hline
SRCC     & 0.7409 & 0.7956 & \textbf{0.8270} & 0.8112 \\
\hline
\hline
\multicolumn{5}{c}{TDN}                       \\\hline
GCN layer & 1      & 2      & 3      & 4      \\\hline
SRCC     & 0.8009 & 0.8014 & \textbf{0.8270}& -    \\
\bottomrule[1.5pt]
\end{tabular}
\label{tab:arch}
\end{table}

For each setting of hyperparameters and architecture, we conduct pre-training process until the loss function converges. We conduct the linear evaluation experiments on validation set, Kadid-10k dataset, as~\cite{he2020momentum}, where the parameters of DGRs generation part (Backbone, Node Builder and Edge Builder) are fixed and the linear convolutional layers are trained to regress the MOS. Then the SRCC results are reported.

\paragraph{Architecture}
In this section, we provide experiments of network architecture on KonIQ dataset~\cite{lin2018koniq}, which is shown in Fig.~\ref{fig:network} and the results are shown in TABLE~\ref{tab:arch}. All the results are tested on models trained on $100,000$ epochs. All the other parameters are kept consistent, and only the parameters to be compared are changed. 
For the Node Builder, we test the performance of $1$, $2$, $3$, and $4$  fully connected (fc) layers. 
It is observed that when Node Builder with $3$ fully connected layers achieves the best performance.
We also test the performance on different number of GCN layers of Edge Builder and TDN. Edge Builder with $3$ graph convolutional layers and TDN with $3$ graph convolutional layers, our model achieves the best performance (when the number of graph convolutional layers in TDN is $4$, the pre-training is unstable).

We also test the relationship between the number of epoch and performance on LIVEC dataset and CSIQ dataset. In the pre-training process, the model's ability of distinguishing and representing each distortion is improved, leading to the performance improvements on both synthetic distortion dataset and authentic distortion dataset. However, long-term training cannot continue to improve the performance, because overfitting to synthetic distortion dataset leads to poor generalization on unknown distortion types. 

\paragraph{Complexity Analysis}

In this section, we analyze the complexity of each module in GraphIQA. We list the parameter amount (noted as Param.) of each module in GraphIQA in TABLE~\ref{tab:complexity}. The total amount is $34.9$M. It is notable that TDN and FPN do not participate in the finetuning and inference stage, which makes the actual amount is $29.9$M.

\begin{table}[h]
\centering
\caption{The number of parameters of modules.}
\begin{tabular}{c|ccccc}
\toprule[1.5pt]
       &Backbone & NB   & EB   & TDN  & FPN  \\ \hline
Param. &23.5M    & 3.7M & 2.7M & 2.8M & 2.2M \\ \bottomrule[1.5pt]
\end{tabular}
\label{tab:complexity}
\end{table}

\begin{table}[h]
\centering
\caption{Linear evaluation results with different margins of triplet loss.}
\begin{tabular}{c|ccccc}
\toprule[1.5pt]
Margin & 0 & 0.2 & 0.5 & 1 & soft margin~\cite{hermans2017defense} \\ \hline
SRCC  & 0.7939 & \textbf{0.8270} & 0.8130 & 0.7464 & 0.7957  \\ 
\bottomrule[1.5pt]
\end{tabular}
\label{tab:margin}
\end{table}



\paragraph{Margin of triplet loss}
The margin setting in the loss function will directly affect how well the network can discriminate the distortion types. The small the margin, the greater the discriminability of the learned DGR. When distance between two graph is smaller than margin the loss is set as $0$. When soft-margin is used, there is no truncation in the loss function, and the distance between similar samples can be as small as possible.
Experiments on different margins of triplet loss are provided in TABLE~\ref{tab:margin}. It is observed that when it is set as $0.2$, the GraphIQA gets the best performance on linear evaluation result.

\section{Conclusion}
\label{conclusion}

In this paper, we integrate graph representation learning into IQA and propose a novel framework GraphIQA to learn DGRs. Having the ability to represent the characteristics of each distortion and the internal structure, GraphIQA can not only generate DGRs as prior knowledge when processing known distortions but also infer the influence of unknown distortions on the perceptual image quality.
For future work, for better distortion representation of distortion, a more complex graph structure can be considered to optimize the existing model, e.g., the integration of hyper-nodes.
To explore richer application scenarios, based on our learned DGR, we will challenge interpretable IQA problems. Besides, noting that DGR can well represent distortions, we will also try to utilize GraphIQA to participate in helping image restoration tasks, such as denoising or deblurring.


\ifCLASSOPTIONcaptionsoff
  \newpage
\fi

\bibliographystyle{IEEEtran.bst}
\bibliography{egbib}




\end{document}